\documentclass[aps,prb,twocolumn,superscriptaddress,showpacs,footinbib]{revtex4}
\usepackage{times}
\usepackage{amsmath}
\usepackage{graphicx}

\begin{document}

\title{Optical transitions in quantum-ring complexes}
\author{T.~Kuroda}
\affiliation{Nanomaterials Laboratory, National Institute for Materials Science,
Namiki 1-1, Tsukuba 305-0044, Japan}
\affiliation{PRESTO, Japan Science and Technology Agency,
4-1-8 Honcho Kawaguchi, Saitama, Japan}
\author{T.~Mano}
\author{T.~Ochiai}
\affiliation{Nanomaterials Laboratory, National Institute for Materials Science,
Namiki 1-1, Tsukuba 305-0044, Japan}
\author{S.~Sanguinetti}
\affiliation{Nanomaterials Laboratory, National Institute for Materials Science,
Namiki 1-1, Tsukuba 305-0044, Japan}
\affiliation{Dipartimento di Scienza dei Materiali, Universit\'{a} di Milano Bicocca, 
Via Cozzi 53, I-20125 Milano, Italy}
\author{K.~Sakoda}
\affiliation{Nanomaterials Laboratory, National Institute for Materials Science,
Namiki 1-1, Tsukuba 305-0044, Japan}
\affiliation{Graduate School of Pure and Applied Science, University of Tsukuba, 
1-1-1 Tennodai, Tsukuba 305-8577, Japan}
\author{G.~Kido}
\affiliation{Nanomaterials Laboratory, National Institute for Materials Science,
Namiki 1-1, Tsukuba 305-0044, Japan}
\affiliation{High Magnetic Field Center, National Institute for Materials Science, 
Sakura 3-13, Tsukuba 305-0003, Japan}
\author{N.~Koguchi}
\affiliation{Nanomaterials Laboratory, National Institute for Materials Science,
Namiki 1-1, Tsukuba 305-0044, Japan}
\date{\today}
\begin{abstract}
Making use of a droplet-epitaxial technique, we realize nanometer-sized quantum ring complexes, 
consisting of a well-defined inner ring and an outer ring. 
Electronic structure inherent in the unique quantum system is analyzed 
using a micro-photoluminescence technique. One advantage of our growth method is 
that it presents the possibility of varying the ring geometry. 
Two samples are prepared and studied: a single-wall ring and a concentric double-ring. 
For both samples, highly efficient photoluminescence emitted from a single quantum structure is detected. 
The spectra show discrete resonance lines, which reflect the quantized nature of the ring-type electronic states. 
In the concentric double--ring, the carrier confinement in the inner ring and that in the outer ring are identified 
distinctly as split lines. The observed spectra are interpreted on the basis of single electron effective mass calculations.
\end{abstract}
\maketitle
\section{Introduction}
Recent progress in nanofabrication technology allows the simulation of novel atomic physical phenomena 
on an artificial platform, such as presence of $\delta$-function-like density of states 
on quantum dots (QD) \cite{GCL95,BGL99}, 
realization of molecular-orbital state on spatially coupled QDs \cite{VGN95,BHH01}, 
and formation of nanometer-sized quantum rings \cite{MCB93,GRS97,MN05,MKS05}, 
which are nanoscopic analogues of benzene. 
Among them, fabrication of semiconductor quantum rings has triggered strong interest 
in realization of quantum topological phenomena, which are expected in a small systems 
with simply-connected geometry \cite{AB59,B84}. 
The Aharonov-Bohm (AB) effect, which engenders so-called \textit{persistent current} \cite{BIL83}, 
has been explored for various types of mesoscopic rings, 
based on metals \cite{metal_rings} and semiconductors \cite{MCB93,semicon_rings}, 
using magnetic and transport experiments. As an optical, \textit{i.e.} \textit{non-contact}, 
approach, 
Lorke \textit{et al.} first observed far-infrared optical response in self-assembled quantum rings, 
revealing a magneto-induced change in the ground state from angular momentum $l = 0$ to $l = -1$, 
with a flux quantum piercing the interior \cite{LLG00}. 
Later, Bayer \textit{et al.} reported pronounced AB-type oscillation 
appearing in the resonance energy of a charged exciton confined in a single lithographic quantum ring \cite{BKH03}; 
furthermore, Ribeiro \textit{et al.} observed the AB signature in type-II quantum dots, 
in which a heavy hole travels around a dot \cite{RGC04}. 

Although magneto-conductance characteristics have garnered considerable attention, optical manifestation of the AB effect has remained a controversial subject \cite{RR00}. 
In this regime, both an electron and a hole are excited simultaneously; the net charge inside a ring decreases to zero. 
Because of the charge neutrality, the loop current associated with a magnetic flux must vanish, 
engendering the absence of the AB effect for a tightly bound electron-hole pair (\textit{exciton}). 
Several theories have been proposed that a composite nature of excitons allows for a non-vanishing AB effect in a small sufficient ring \cite{RR00}, 
and in excited state emissions \cite{HZL01}. 
On the other hand, a negative prediction has also been reported in which the AB effect 
can hardly take place in more-realistic rings with finite width \cite{SU01}. 
A large difference in trajectories for an electron and a hole is necessary to exhibit the AB effect on neutral excitation \cite{GUK02}. 
\begin{figure}
\includegraphics[width=7.5cm,clip]{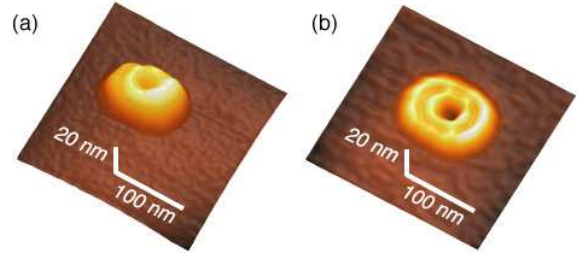}
\caption{\label{fig1}(color online) Atomic force microscope images in 250 $\times$ 250~nm$^2$ area for (a) a GaAs quantum ring (QR), and (b) a concentric double ring (DQR), grown by droplet epitaxy. After ring formation, they are covered by an Al$_{0.3}$Ga$_{0.7}$As layer for optical experimentation.}
\end{figure}

The present study examines the optical transition in \textit{strain-free} self-assembled GaAs quantum rings 
in a zero magnetic field. We have recently reported self-production of nanometer-sized GaAs rings on (Al,Ga)As 
by means of a droplet epitaxial technique \cite{MN05,MKS05}. 
Because of their good cyclic symmetry, together with high tunability of the ring size and shape, 
the present system is expected to open a new route to implement the AB effect within the optical regime. 
Both the ground-state transition and the excited-state transition are identified 
by single quantum ring photoluminescence (PL). 
The spectra are found to be in good agreement with results of single-carrier calculation. 

The paper is organized as follows. In Sec. II, we briefly explain the sample preparation 
and the procedure of the optical experiment. In Sec. III, we present low temperature PL spectroscopy 
of single quantum rings. Section IV presents the theoretical results; we discuss the experimental data 
in terms of this model in Sec. V. Our conclusions are summarized briefly in Sec. VI. 

\section{Experimental Procedure}
\subsection{sample preparation} 
GaAs quantum rings were grown on Al$_{0.3}$Ga$_{0.7}$As using modified droplet epitaxy \cite{KTC91}. 
In this growth, cation (Ga) atoms are supplied solely in the initial stage of growth, producing nanometer-sized droplets 
of Ga clusters. After formation of the Ga droplets, anion atoms (As) are supplied, 
leading to crystallization of the droplets into GaAs nanocrystals. 
In contrast to the other methods to fabricate QDs, such as Stranski-Krastanow growth, 
this technique can produce \textit{strain-free} quantum dots based on lattice-matched heterosystems. 
In addition to these characteristics, we recently found that it has a high controllability of the crystalline shape: 
When we irradiate the Ga droplets with an As beam of sufficiently high intensity, 
typically $2 \times 10^{-4}$ Torr beam equivalent pressure (BEP) at 200 $^{\circ}$C, 
the crystalline shape becomes cone-like, following the shape of the original droplet \cite{WKG00,WTGK01}. 
When we reduce the As intensity to $1 \times 10^{-5}$ Torr BEP, the QD becomes ring-like, 
with a well-defined center hole \cite{MN05}. Further reduction of the As flux, down to $2 \times 10^{-6}$ Torr BEP, 
produces the striking formation of unique concentric double-rings: an inner ring and an outer one \cite{MKS05}. 
The rings show a good circular symmetry, whereas small elongation is found along the [0-11] direction 
(5\% for the inner ring and 8\% for the outer ring). 

Two samples are used in the experiment: a GaAs quantum ring of 40 nm diameter with 15 nm height, (abbreviated to QR, hereafter) and concentric double-rings consisting of an inner ring of 40 nm diameter and 6 nm height, and an outer ring of 80 nm diameter with 5 nm height (abbreviated to DQR). In the growth of these two rings, the same conditions were applied to the initial deposition of Ga droplets (1.75 monolayer (ML) of Ga at 0.05 ML/s to the surface of a Al$_{0.3}$Ga$_{0.7}$As substrate at 300~$^{\circ}$C). Thus, the mean volume for each structure is expected to be equivalent for QR and DQR, whereas its crystalline shape differs drastically. After ring formation, they were capped by an Al$_{0.3}$Ga$_{0.7}$As barrier of 100 nm thickness, following rapid thermal annealing (RTA; 750 $^{\circ}$C for 4 min). The ring shape before capping is characterized by atomic force microscopy (AFM), as shown in Fig.~\ref{fig1} (see also the averaged cross-section of DQR presented in Fig.~\ref{eh_level}(c)). The ring density is $1.3 \times 10^{8}$ cm$^{-2}$ for both samples, allowing the capture of the emission from a single structure using a micro-objective setup. 

We would like to stress the difference in growth process between these quantum rings and  In(Ga)As rings, whose growth was previously reported \cite{GRS97,GG03}. The ring formation of the latter case is associated with partial capping of a thin GaAs layer on InAs QDs, which were originally made by the Stranski-Krastanow method. Subsequent annealing results in the morphological change from island-like QDs to ring-like nanocrystals. In contrast, the present rings are formed at the crystalline stage of GaAs. The ring shape in this case is determined by the flux intensity of As beams. After the formation of rings, they are capped by a thick (Al,Ga)As layer. Later we apply RTA to improve their optical characteristics. Note that the final RTA processing does not modify the nanocrystalline ring shape, according to the negligible interdiffusion of Ga and As at a GaAs/(Al,Ga)As heterointerface at the relevant temperature. \cite{SK86}
\begin{figure}
\includegraphics[width=6.5cm]{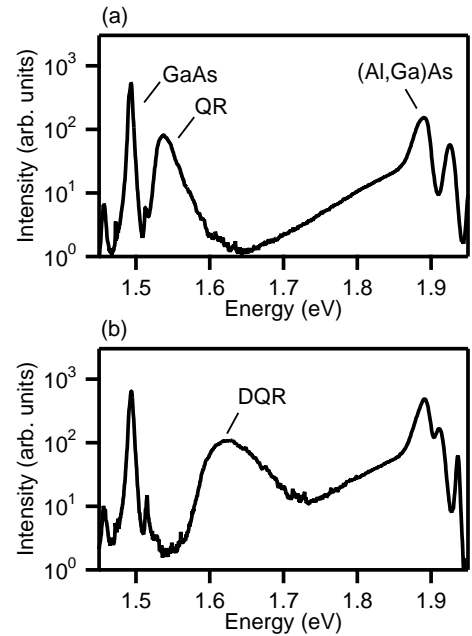}
\caption{\label{macroPL}Far-field emission spectra of the sample with (a) QRs and (b) DQRs at 5 K plotted on a logarithmic scale. The excitation density is 50 mW/cm$^2$.}
\end{figure}

\subsection{optical arrangement}
In the PL experiment, we used a continuous wave He-Ne laser as an excitation source. The laser emitted 544 nm wavelength light, corresponding to 2.28 eV in energy. The excitation beam from the laser was obliquely incident to the sample. It was loosely focused by a conventional lens (30-cm focal length) into an elliptical spot of $0.5 \times 1.4$ mm$^{2}$. Emission from the sample was collected by an aberration-corrected objective lens of N.A. (numerical aperture) = 0.55. Combination of the objective lens and a pinhole (50-$\mu$m diameter) at the focal plane allowed the capture of emissions inside a small spot of 1.2-$\mu$m diameter. For this spot size, $1.3$ rings were expected to lie in the focus on average. The position of detection was translated laterally so that single quantum structures were captured individually. For this purpose, we moved the objective lens with sub-micrometer precision using piezo transducers. During translation of the spot, the condition of excitation was kept unchanged because the excitation area was sufficiently larger than the area that covered the entire translation. The emission passing through the pinhole was introduced into an entrance slit of a polychromator of 32-cm focal length. After being spectrally dispersed, it was recorded by a charged-couple device detector with 0.8-meV resolution. In advance of the micro-objective measurement, we used a conventional PL setup to observe the spectra of the ring ensemble. All experiments were performed at 3.8 K. 

Before describing experimental results, we mention the carrier dynamics associated with photoexcitation of the present condition. For our laser beam, photocarriers are produced initially in the barrier, whose band gap is 1.95 eV. After diffusion, the carriers are captured by the quantum rings. The captured carriers then relax into the lower lying quantum ring levels where they radiatively recombine. Our previous study showed the recombination lifetime in GaAs/(Al,Ga)As QDs as $\sim$ 400 ps, whereas the characteristic time of intra-dot relaxation was much shorter -- less than 30 ps -- depending on excitation density \cite{SWT02, KSG02}. Because of the rapid intra-dot relaxation, we can expect that an electron and a hole recombine after they are in quasi-equilibrium. The quantized levels are occupied by carriers according to the Fermi distribution. Similar dynamics are expected in the ring system. 

\section{Experimental Results}
\subsection{PL from the ensemble of quantum rings}
We present the far-field PL spectrum of the sample with QR in Fig. \ref{macroPL}(a). It comprises several spectral components. The sharp line at 1.49 eV is assigned by impurity-related emissions from the GaAs substrate. Because of a thin deposited layer of the sample (500 nm thickness), the excitation beam is expected to reach the substrate, thereby producing strong emissions. The broad emission band at 1.544 eV in the center energy is attributed to recombination of an electron and a hole, which are confined in GaAs QRs. The spectrum is broadened by 28 meV in full-width-at-half-maximum (FWHM). The line broadening is caused by the size distribution of the rings. Several emission components ranging from 1.7 eV to 1.95 eV are assigned by recombinations in the (Al,Ga)As barrier. The spectral tail to lower energy suggests the presence of impurities and imperfections in the barrier layer. Note that the excitation density at this measurement is quite low (50 mW/cm$^2$). For that reason, the impurity-related signal should be relatively emphasized. 

The PL spectrum of the sample with DQR is presented in Fig. \ref{macroPL}(b). The signals that are related to the GaAs substrate and to the barrier are identical to those of QR. The emission band at 1.628 eV in center energy originates from recombination of the ensemble of DQRs. It is broadened by 49 meV in FWHM. We find that the PL energy of DQR is higher than that of QR. The energy shift reflects the small height of DQR. In our rings, stronger confinement is induced along the growth direction than in the lateral in-plane direction. Thus, the reduction in height, associated with formation of DQR, enhances their confinement energy, causing a blue shift in the PL spectrum.

\subsection{Spectroscopy of a single quantum ring (QR)}
\begin{figure}
\includegraphics[width=5.5cm]{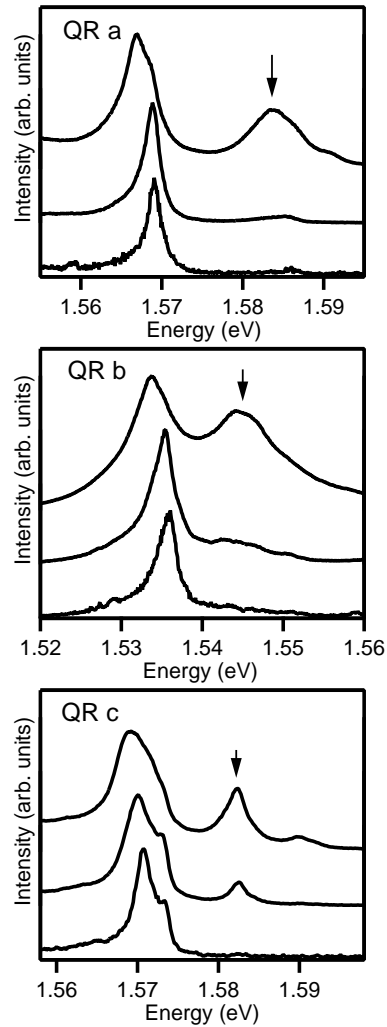}
\caption{\label{fig_QR}Emission spectra for a single GaAs QR. Three examples -- a, b, and c -- are presented. Their respective excitation densities were, from bottom to top, 1, 10, and 30 W/cm$^2$. Spectra are normalized to their maxima and offset for clarity.}
\end{figure}
In Fig.~\ref{fig_QR} we show the PL spectra of three different quantum rings, QR a, -b, and -c, and their dependence on excitation intensity. In QR a at low excitation, we find a single emission line appearing at 1.569 eV, which results from recombination of an electron and a hole, both occupying the ground state of the ring. With increasing excitation intensity, a new emission line, indicated by an arrow, emerges at 1.582 eV. Further increase in excitation density causes saturation in the intensity of the original line along with a nonlinear increase in the new line. Superlinear dependence of the emission intensity suggests that the satellite line comes from the electron--hole recombination from an excited level of the ring. Thus, the energy difference between the the ground and the excited state in the QR is 13 meV. 

In addition to the state-filling feature associated with photoinjection, we find the ground-state emission being shifted to low energy. It is a signature of multi-carrier effects. In the presence of multiple carriers inside a ring, their energy levels are modified by the Coulomb interaction among carriers. Because the optical transition energy is mainly renormalized according to the exchange correction, the many-carrier effects results in spectral red shift of the emission, depending on the number of carriers. Similar features have been observed in numerous quantum dot systems including GaAs dots \cite{KSG02} and In(Ga)As rings \cite{WSH00}. Note that the relevant multiplet is not spectrally resolved in our rings, but it leads to the red shift of the broad emission spectra. Moreover, the biexcitonic emission is expected to contribute to the low energy tail of the spectra because the biexciton binding energy was found to be $\sim$0.8 meV in our droplet-epitaxial GaAs dots \cite{KSG02,KKS05}. At high excitation, we also find spectral broadening, which is attributed to an incoherent collision process that occurs among carriers. 

Identical spectral features are found in QR b. The ground-state emission is observed at 1.537 eV, whereas the excited-state emission appears at 1.546 eV. The energy spacing between these two lines is 9 meV. Although dependence on excitation density is quite similar for QR a and QR b, the energy levels are slightly different because of a small dispersion of the ring shape and size. We note that the linewidth of the ground state emission is also different between QR a (2.2 meV FWHM) and QR b (3.4 meV). Moreover, these are considerably large compared with that known for self-assembled QDs. We ascribe the line broadening to spectral diffusion, \textit{i.e.}, an effect of the local environment that surrounds each QR: Our samples are expected to contain a relatively large density of imperfections and excess dopants, associated with the low-temperature growth of the sample. It causes low-frequency fluctuation in the local field surrounding QRs, which is due to the carrier hopping inside the barrier. This leads to efficient broadening of the PL spectra, which depend on the local environment of each ring. Detailed examination of the origin of line broadening is studied in droplet-epitaxial GaAs quantum dots \cite{KKS05}. 

In QR c, we find a shoulder structure on the ground-state emission, which suggests the split in the relevant level caused by lateral elongation, and/or structural imperfection inherent in this ring. Broken symmetry in the ring shape causes degeneracy lift in the energy level, leading to observation of the doublet spectra. \cite{footnote1} Apart from this split structure, the same spectral characteristics are found in QR c. The energy difference from the original line to the satellite is observed to 11 meV, which is between the value of QR~a and that of QR~b.

\begin{figure}
\includegraphics[width=8cm]{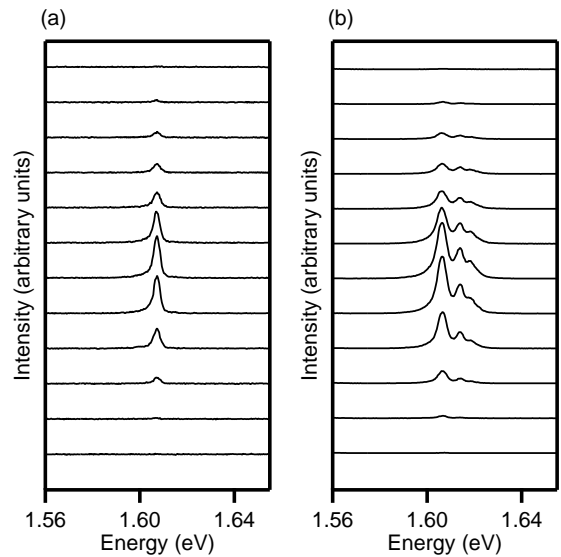}
\caption{\label{fig_spatial}Position dependence of micro PL spectra in GaAs QR at (a) 1.2 W/cm$^{2}$ and (b) 36 W/cm$^{2}$. From top to bottom the position of detection is moved from 0 to 3.8 $\mu$m in steps of 0.32 $\mu$m. Spectra are vertically offset for clarity.}
\end{figure}

Spatial dependence of the PL spectrum is shown in Fig.~\ref{fig_spatial}, where the position of detection is laterally translated on the sample in steps of 0.32 $\mu$m. Figure \ref{fig_spatial}(a) presents the results obtained at low excitation. They exhibit a single line associated with the ground-state emission from a single QR, depending on the position of detection. Lateral broadening of the emission is estimated as $\sim$1.2 $\mu$m FWHM, which is consistent with the spatial resolution of our micro-objective setup. At high excitation, the spectra change into multiplets, as shown in Fig.~\ref{fig_spatial}(b). They show the same lateral profile with those obtained at low excitation, confirming the multiplet being emitted from a single QR, and not from multiple QRs with different energies. The energy split from the ground-state emission to the first-excited-state emission is observed to be 9 meV for this QR.

\subsection{spectroscopy on a single concentric double-rings (DQR)}
\begin{figure}
\includegraphics[width=5.5cm]{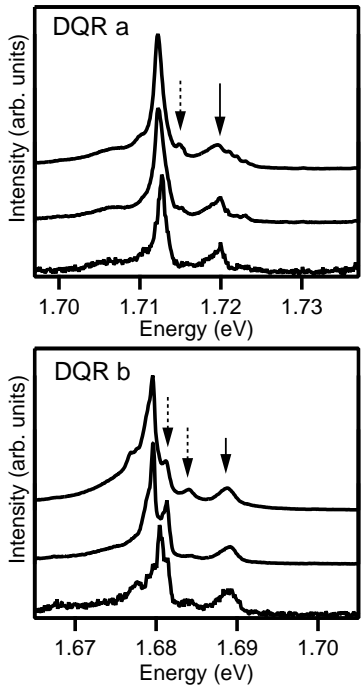}
\caption{\label{fig_DQR}Emission spectra for concentric quantum double-rings, DQR a and -b. Their respective excitation densities were, from bottom to top, 1, 10, and 30 W/cm$^2$. Spectra are normalized to their maxima and offset for clarity. }
\end{figure}
We present the PL spectra of two concentric quantum double-rings -- DQR a and DQR b -- in Fig.~\ref{fig_DQR}. Similarly to the case of QR, the spectra consist of discrete lines, \textit{i.e.}, a main peak following a satellite one, which is at the high energy side of the main peak. The former is associated with recombination of carriers in the ground state, whereas the latter is from the excited states. The energy difference between the ground-state line and the excited-state one is 7.2 meV in DQR a and 8.5 meV in DQR b. We point out that, in contrast to the QR case, we observe the satellite peak even at the lowest excitation. For the lowest excitation intensity, we can estimate the carrier population inside a ring to be less than 0.1, according to the carrier capturing cross-section determined for GaAs QD, which was prepared with the same epitaxial technique \cite{KSG02}. Observation of the excited state emission suggests reduction of carrier relaxation from the excited level to the ground level. That feature will be discussed later.

At high excitation, we find that several additional lines are superimposed on the spectra, as shown by the broken arrows. Presence of these contributions implies the presence of fine energy structures in DQR. 

\section{Calculation for single-carrier levels}
\begin{figure*}
\includegraphics[width=15cm]{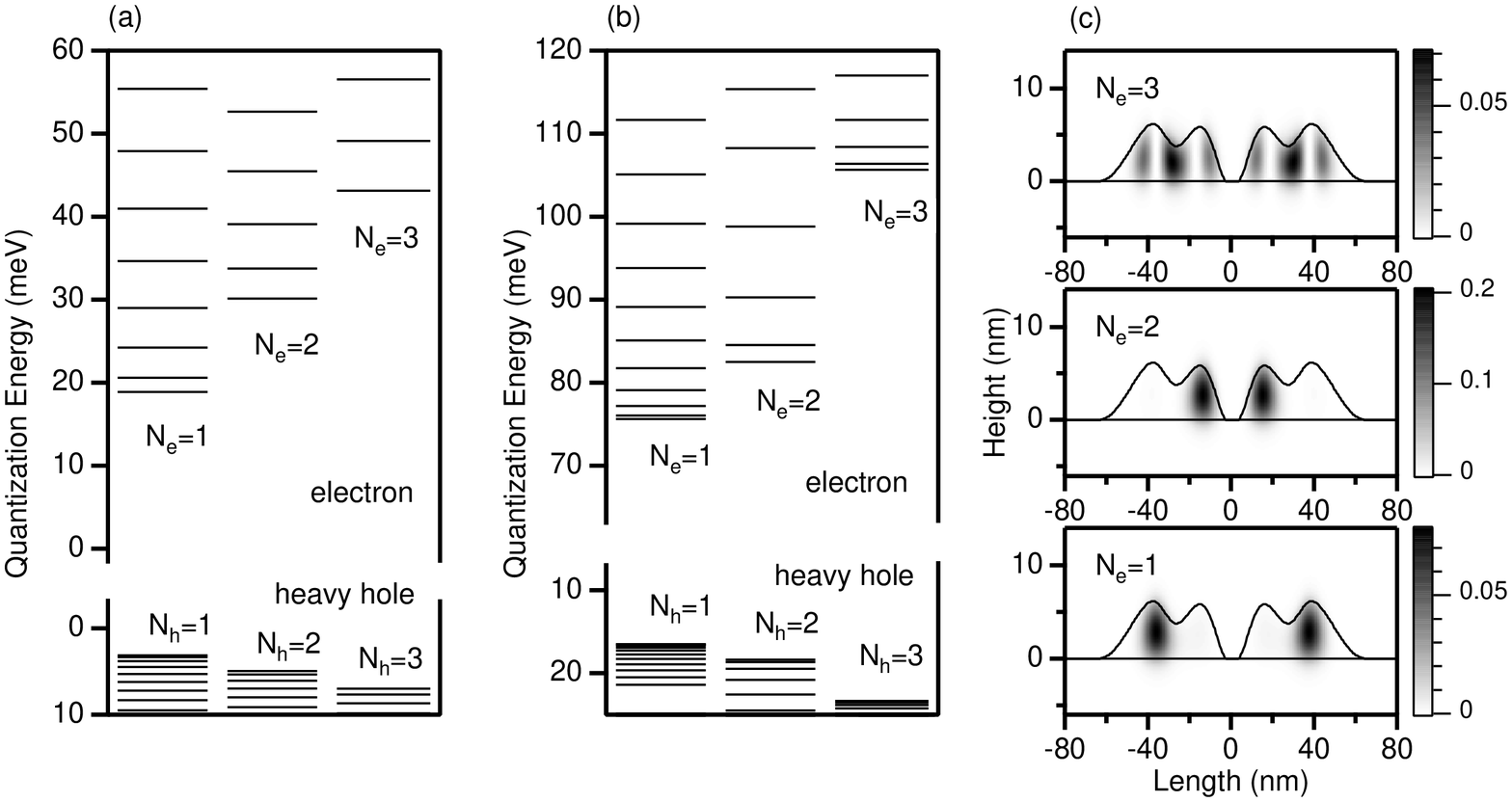}
\caption{\label{eh_level}Single-carrier energy levels in (a) QR and (b) DQR. Quantization energies for an electron (a heavy hole) with the three lowest radial quantum numbers, $N_{e(h)}$, and various angular momenta (up to 10) are presented. (c) Cross-sectional imaging of electronic probability density in DQR for $N_{e} =1, 2, \text{ and } 3$ with $L = 0$. The line represents the potential of confinement used for calculation.}
\end{figure*}
\begin{table}
\caption{\label{table1}Material parameters used in the effective mass calculation for the conduction band (CB) and the valence band (VB)}
\begin{ruledtabular}
\begin{tabular}{lccc}
Quantity & Units & GaAs & Al$_{0.3}$Ga$_{0.7}$As \\ \hline
CB effective mass \footnotemark[1] & \textit{m}$_0$ & 0.067 & 0.093 \\
VB effective mass (heavy hole) \footnotemark[1] & \textit{m}$_0$ & 0.51 & 0.57 \\ \hline
CB band offset \footnotemark[2] & meV & \multicolumn{2}{c}{262} \\
VB band offset \footnotemark[2] & meV & \multicolumn{2}{c}{195} \\
\end{tabular}
\end{ruledtabular}
\footnotetext[1]{Reference \protect\onlinecite{PG94}.}
\footnotetext[2]{Reference \protect\onlinecite{YSM04}.}
\end{table}
We evaluate the energy levels of the ring in the framework of a single-band effective-mass envelope model \cite{MB94,CH00}. In calculation, the actual shape measured by AFM is adopted as the potential of quantum confinement; for simplicity, the ring is assumed to hold a cylindrical symmetry. The technique employed in this section follows Ref.~\onlinecite{MB94} describing the exact diagonalization of the effective-mass Hamiltonian. Note that, in our \textit{lattice-matched} GaAs/(Al,Ga)As rings, strain effects are negligible. For that reason, the simple effective-mass approach is expected to provide accurate energy levels. This presents a contrast to the case of Stranski-Krastanow grown dots, where the electronic structure is modified strongly by complex strain effects \cite{SGB99}. The versatility of the present method is seen in Refs.~\onlinecite{SWK02} and \onlinecite{MSG04}, showing good agreement between the asymmetric PL lineshape in a GaAs/(Al,Ga)As QD ensemble and the calculation, taking into account the morphologic distribution of dots. 

We also notice that the present calculation neglects Coulomb interaction between an electron and a hole. Because our rings are sufficiently small that confinement effects are dominant, the Coulomb interaction can be treated as a constant shift in the transition energies, independent of the choice of an electron state and the hole state. In following discussion, we are interested in relative energy shifts from the ground-state transition to the excited-state one, and dependence of the exciton binding energy on relevant (single-carrier) transition should be sufficiently below the experimental accuracy. Thus, we restrict ourselves to calculate the single carrier energy levels, discarding Coulomb correlation effects. 

Our approach to the problem is to enclose the nanocrystal inside a large cylinder of radius $R_{c}$ and height $Z_{c}$, 
on the surface of which the wavefunction vanishes. 
Care should be taken to set $R_c$ and $Z_c$ away from the ring, 
so that the eigen values are almost independent of their choice. 
Taking into account the rotational symmetry of the Hamiltonian, and for this boundary condition, 
the wave function, $\Phi_L$, where $L(=0, \pm1, \cdots)$ is the azimuthal quantum number, 
is expanded in terms of a complete set of the base functions, 
$\xi ^{L}_{i,j}$, formed by products of Bessel functions of integer order $L$ and sine functions of $z$, 

\begin{eqnarray}
\Phi _L (z,r,\theta ) = \sum_{i,j>0} A^{L}_{i,j} \xi ^{L} _{i,j}(z,r,\theta ),\\
\xi ^{L} _{i,j}(z,r,\theta) = \beta ^{L}_{i} J_{L}(k^L_{i}r) e^{iL\theta} \sin (K_{j}z),
\end{eqnarray}
where $k^L_i R_c$ is the $i$ th zero of the Bessel function of integer order $J_{L}(x)$, $K_j = \pi j / Z_c$, and $\beta ^L _i$ is appropriate normalization factors, \textit{i.e.}, 
\begin{equation}
\beta ^L_i = \sqrt{ \frac {2}{\pi Z_{c} R_{c}^{2}}} \frac{2}{|J_{L-1}(k^{L}_{i}R_c)-J_{L+1}(k^{L}_{i}R_c)|}.
\end{equation}
In advance of calculation, we have prepared a Hamiltonian that includes a potential term in cylindrical coordinates. Then, we calculated its matrix elements through numerical integration with $\xi ^{L} _{i,j}$ over $r$ and $z$. Finally, the eigen states were obtained with diagonalizing matrix. For the present calculation, we have taken into account 35 Bessels and 35 sine functions as the base functions for each value of $L$, and $R_{c} (Z_{c}) = 120 (20)$ nm. Material parameters used in calculation are summarized in Table~\ref{table1}. 

A series of single carrier levels of QR is shown in Fig.~\ref{eh_level}(a). Because the system has cylindrical symmetry, each carrier level is specified by the principal (\textit{radial}) quantum number \cite{footnote2} $N (=1, 2, \cdots)$, and an azimuthal quantum number $L$, corresponding to the angular momentum. Two levels with $\pm L$ are degenerated at zero magnetic field. In Fig.~\ref{eh_level}(a), the carrier levels belonging to each radial quantum number are aligned vertically with those of a different angular momentum. We find that a vertical ($L$-dependent) sequence of quantized levels shows a typical signature of ring-type confinement. For an ideal ring with infinitesimal width, being treated as a one-dimensional system with translational periodicity, the level series is expressed as 
\begin{equation}
E_{L}=\frac{\hbar ^2}{2m^{*}R^2}L^2, 
\end{equation}
where $R$ and $m^{*}$ respectively represent the radius of the ring and the carrier mass. We find that the bilinear dependence of the level series, shown in Eq.~(4), is reflected clearly in the line sequence in Fig.~\ref{eh_level}(a). 

The energy levels in DQR are shown in Fig.~\ref{eh_level}(b). As a result of the smaller height, the quantization energies are larger than those of QR. The level sequence of $N =1$ is more densely populated than that of $N = 2$, suggesting a large difference in carrier trajectories between the two levels. According to Eq.~(4), the situation corresponds to the large effective value of $R$ for $N =1$. The fact is confirmed by the wavefunctions shown in Fig.~\ref{eh_level}(c). This figure illustrates the envelope wavefunction of an electron with various values of $N$. They are of zero angular momentum. We find that the electron of $N = 1$ is confined mainly in the outer ring. That of $N = 2$ is in the inner ring, and that of $N = 3$ is situated in both rings. That differential confinement engenders remarkable changes in their trajectory. The amount of penetration for the electron of $N=1$ to the inner ring is found to be $\sim$0.1, whereas that of $N=2$ to the outer ring is $\sim$0.05. 

\section{Discussion}
We have evaluated the oscillator strength for transitions between each electron-hole (\textit{e-h}) level to determine a consistency between the emission spectra and the theoretical levels. The magnitude is proportional to the overlap of corresponding (envelope) wavefunctions. Note that, because of cylindrical symmetry, optical transition is not allowed for an electron and a hole with different angular momentum. Moreover, we have determined the transition strengths for the \textit{e-h} pair with different $N$s as less than 1/10 smaller than that with the same $N$s. We can therefore infer that the electron, specified by a pair of $N$ and $L$, recombines only with the hole of the same $N$ and $L$. This selection rule allows us to describe each optical transition simply by $(N, L)$. 

\begin{figure}
\includegraphics[width=5.5cm]{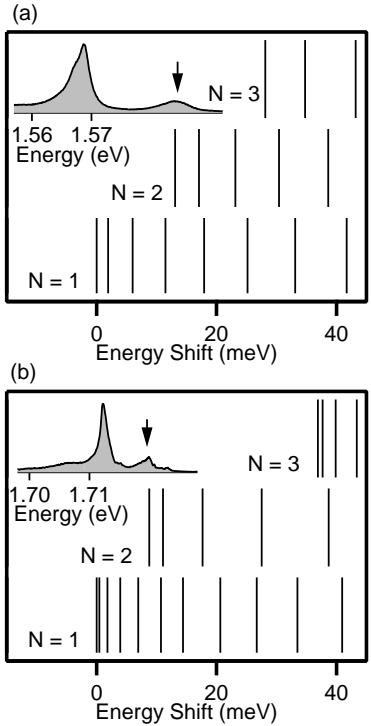}
\caption{\label{pl_vs_cal} (a) A series of optical transition energies in QR, obtained by the calculation. The PL spectrum of QR a at 15 W/cm$^2$ is shown in the inset. (b) The energies of optical transitions in DQR, together with the PL spectrum of DQR a at 10 W/cm$^2$ for comparison.}
\end{figure}

A series of transition energies for QR, obtained by the procedure described above, are shown in Fig.~\ref{pl_vs_cal}(a). For comparison, the emission spectra of QR~a are plotted as an example of experimental data. The main peak and the high-energy satellite in the observed spectra are assigned respectively by the recombination of the \textit{e-h} pair in the lowest state, $(N, L)=(1, 0)$, and that of the first excited \textit{radial} state, $(2, 0)$. The split between the two transitions is deduced to be 13.1 meV, which agrees with the energy shift obtained by experiments. Note that the emissions associated with high angular momenta are not present in the data, which suggests rapid relaxation of angular momentum, whose process is quite faster than transition between radial quantization levels or recombination between an electron and a hole. A possible origin for fast angular momentum relaxation is structural asymmetry of the ring, which results from elongation, impurity, and surface roughness. In this case, angular momentum does not represent a good quantum number, and scattering between different $L$ levels efficiently occurs. 

We show the comparison between the experimental spectra of DQR and the results of calculation in Fig. \ref{pl_vs_cal}(b). As in the case of QR, the main PL peak and the satellite one are explained respectively in terms of the transition of $(N,L)=(1,0)$ and that of $(2,0)$. The energy split deduced from calculation is 8.8 meV, which agrees with the experimental value. It is noteworthy that, in DQR, the wavefunction of $N=1$ is localized mainly in the outer ring, whereas that of $N=2$ is localized in the inner ring. Thus, the two peaks in the observed spectra come from the two rings, which consist of a DQR. In this connection, it is noteworthy that the excited-state emission in our experiment appears even when the carrier population is less than one. This presence of the excited state emission constitutes direct evidence for the carrier confinement into the two rings. Tunneling probability between the inner ring and the outer one is not very large, engendering the observation of the excited-state emission. Note also that the emission of the outer ring is more intense than that of the inner ring. The effect is attributable to their different surface areas, which affects the efficiency of the carrier injection from a barrier. 

Finally, we would like to discuss the validity of the theoretical treatment which neglects the Coulomb correlation between an electron and a hole. A limitation of the validity takes place when an exciton binding energy is fairly large compared to a single-carrier split energy, i.e., 13.1 meV for QR, and 8.8 meV for DQR. In this case the Coulomb correlation admixes various single-carrier levels. We can roughly evaluate the exciton binding energy of the rings in comparison with that of GaAs/(Al,Ga)As quantum wells (QWs). This is because the carrier quantization of our rings is mainly associated with vertical confinement, and the lateral dimension is larger than the exciton Bohr diameter. According to numerous attempts on the study of GaAs/(Al,Ga)As QWs \cite{MK85}, the QW confinement enhances the exciton binding energy from a bulk value of 3.7 meV to $\lesssim$10 meV at $\sim$6-nm thick QWs. A smaller thickness results in weaker exciton binding due to the carrier penetration to a barrier. These results support the exciton binding energy being smaller than or at most comparable to the single-carrier split energy of the rings. The observed spectral doublet, therefore, directly reflects the single-carrier levels. The situation presents a clear contrast to that of Stranski-Krastanow grown QDs, in which the dot dimension is quite smaller than our droplet-epitaxial nanostructures, so that the exciton binding energy reaches $\sim$ 30 meV.

\section{Conclusions}
We have used a remarkable change in quantum dot shape through droplet epitaxial growth to fabricate unique semiconductor quantum ring complexes. Electronic structures of the quantum rings are identified using an optical, non-contact approach. In the small ring-like system, carriers are quantized along two orthogonal degrees of freedom -- radial motion and rotational motion. The latter corresponds to angular momentum. The optical transition takes place on recombination of an electron and a heavy hole, which are in the grand state of the ring, and in the excited radial state. In concentric double quantum rings, emission originating from the outer ring and that from the inner ring are observed distinctly. Results of effective-mass calculations well reproduce the emission spectra applied to a single quantum ring. 

We believe that the present ring system will contribute to a deeper understanding of quantum interference effects in a non-simply connected geometry. In this connection, we would like to point out that our concentric double-rings are a good candidate to realize in-plane polarization for carriers, producing a robust Aharonov-Bohm feature in neutral excitonic transition \cite{GUK02}. Magnet-optical experiments using these quantum rings are now in progress. 

\begin{acknowledgments}
We are grateful to Drs. J. S. Kim, T. Noda, K. Kuroda, and Prof. M. Kawabe for their fruitful discussions. We would like to thank K. Kurakami for his experimental assistance. T.~K. acknowledges a support of a Grant-in-Aid from the Ministry of Education, Culture, Sports, Science and Technology (15710076).
\end{acknowledgments}

\bibliographystyle{apsrev}

\begin{thebibliography}{99}

\bibitem{GCL95}
	M. Grundmann, J. Christen, N. N. Ledentsov, J. B\"{o}hrer, D. Bimberg, 
	S. S. Ruvimov, P. Werner, U. Richter, U. G\"{o}sele, J. Heydenreich, 
	V. M. Ustinov, A. Yu. Egorov, A. E. Zhukov, P. S. Kop'ev, and Zh. I. Alferov, 
	Phys. Rev. Lett. \textbf{74}, 4043 (1995). 

\bibitem{BGL99}
	D.~Bimberg, M.~Grundman, and N.~N.~Ledentsov, 
	\textit{Quantum Dot Heterostructures} 
	(John Wiley \& Sons, Chichester, 1999). 
	
\bibitem{VGN95}
	N. C. van der Vaart, S. F. Godijn, Y. V. Nazarov, C. J. P. M. Harmans, 
	J. E. Mooij, L. W. Molenkamp, and C. T. Foxon, 
	Phys. Rev. Lett. \textbf{74}, 4702 (1995). 
	
\bibitem{BHH01}
	M.~Bayer, P.~Hawrylak, K.~Hinzer, S.~Fafard, M.~Korkusinski, Z.~R.~Wasilewski, 
	O.~Stern, and A.~Forchel, 
	Science \textbf{291}, 5503 (2001).

\bibitem{MCB93}
	D. Mailly, C. Chapelier, and A. Benoit, 
	Phys. Rev. Lett. \textbf{70}, 2020 (1993).
	
\bibitem{GRS97}
	J.~M.~Garc\'{i}a, G.~Medeiros-Ribeiro, K.~Schmidt, T.~Ngo, J.~L.~Feng, 
	A.~Lorke, J.~Kotthaus, and P. M. Petroff, 
	Appl. Phys. Lett. \textbf{71}, 2014 (1997).
	
\bibitem{MN05}
	T. Mano and N. Koguchi, 
	J. Crystal Growth, \textit{in press}.

\bibitem{MKS05}
	T. Mano, T. Kuroda, S. Sanguinetti, T. Ochiai, T. Tateno, J. Kim, T. Noda, 
	M. Kawabe, K. Sakoda, G. Kido, and N. Koguchi, 
	Nano Letters \textbf{5}, 425 (2005).
	
\bibitem{AB59}
	Y. Aharonov and D. Bohm, 
	Phys. Rev. \textbf{115}, 485 (1959). 

\bibitem{B84}
	M. V. Berry, 
	Proc. R. Soc. Lond. A \textbf{392}, 45 (1984).

\bibitem{BIL83}
	M. B\"{u}ttiker, Y. Imry, and R. Landauer, 
	Phys. Lett. \textbf{96A}, 365 (1983). 

\bibitem{metal_rings}
	R. A. Webb, S. Washburn, C. P. Umbach, and R. B. Laibowitz, 
	Phys. Rev. Lett. \textbf{54}, 2696 (1985); 
	L. P. L\'{e}vy, G. Dolan, J. Dunsmuir, and H. Bouchiat, 
	\textit{ibid.} \textbf{64}, 2074 (1990); 
	V. Chandrasekhar, R. A. Webb, M. J. Brady, M. B. Ketchen, W. J. Gallagher, 
	and A. Kleinsasser, 
	\textit{ibid.} \textbf{67}, 3578 (1991); 
	A. Yacoby, M. Heiblum, D. Mahalu, and H. Shtrikman, 
	\textit{ibid.} \textbf{74}, 4047 (1995).
	 
\bibitem{semicon_rings}	
	A. Fuhrer, S. L\"{uscher}, T. Ihn, T. Heinzel, K. Ensslin, W. Wegscheider, 
	and M. Bichler, 
	Nature \textbf{413}, 822 (2001). 

\bibitem{LLG00}
	A.~Lorke, R.~J.~Luyken, A.~O.~Govorov, J.~P.~Kotthaus, 
	J.~M.~Garcia, and P.~M.~Petroff, 
	Phys. Rev. Lett. \textbf{84}, 2223 (2000). 
	
\bibitem{BKH03}
	M. Bayer, M. Korkusinski, P. Hawrylak, T. Gutbrod, M. Michel, 
	and A. Forchel, 
	Phys. Rev. Lett. \textbf{90}, 186801 (2003).
	
\bibitem{RGC04}
	E. Ribeiro, A. O. Govorov, W. Carvalho, Jr., and G. Medeiros-Ribeiro, 
	Phys. Rev. Lett. \textbf{92}, 126402 (2004).

\bibitem{RR00}
	R. A. R\"{o}mer and M. E. Raikh, 
	Phys. Rev. B \textbf{62}, 7045 (2000). 

\bibitem{HZL01}
	H.~Hu, J.~L.~Zhu, D.~J.~Li, and J.~J.~Xiong, 
	Phys. Rev. B \textbf{63}, 195307 (2001). 

\bibitem{SU01}
	J. Song and S.~E.~Ulloa, 
	Phys. Rev. B \textbf{63}, 125302 (2001). 

\bibitem{GUK02}
	A. O. Govorov, S.~E.~Ulloa, K. Karrai, and R.~J.~Warburton, 
	Phys. Rev. B \textbf{66}, 081309(R) (2002). 
	
\bibitem{KTC91}
	N.~Koguchi, S.~Takahashi, and T.~Chikyow, 
	J. Cryst. Growth \textbf{111}, 688 (1991). 

\bibitem{WKG00} 
	K.~Watanabe, N.~Koguchi, and Y.~Gotoh, 
	Jpn. J. Appl. Phys. \textbf{39}, L79 (2000). 
	
\bibitem{WTGK01} 
	K.~Watanabe, S. Tsukamoto, Y. Gotoh, and N.~Koguchi, 
	J.~Cryst.~Growth \textbf{227-228}, 1073 (2001). 
	
\bibitem{GG03}
	D. Granados and J.~M.~Garc\'{i}a, 
	Appl. Phys. Lett. \textbf{82}, 2401(2003).

\bibitem{SK86}
	T. E. Schlesinger and T, Kuech, 
	Appl. Phys. Lett. \textbf{49}, 519(1986).

\bibitem{SWT02} 
	S. Sanguinetti, K.~Watanabe, T. Tateno, M. Wakaki, N.~Koguchi, 
	T. Kuroda, F. Minami, and M. Gurioli, 
	Appl. Phys. Lett. \textbf{81}, 613 (2002).
	
 \bibitem{KSG02}
 	T. Kuroda, S. Sanguinetti, M. Gurioli, K. Watanabe, F. Minami, and N. Koguchi,
	Phys. Rev. B \textbf{66}, 121302(R) (2002).
	
\bibitem{WSH00}
	R. J. Warburton, C. Sch\"{a}flein, D. Haft, F. Bickel, A.~Lorke, 
	K.~Karrai, J.~M.~Garcia, W. Schoenfeld, and P. M. Petroff, 
	Nature \textbf{405}, 926 (2000). 

\bibitem{KKS05}
	K. Kuroda, T. Kuroda, K. Sakoda, N. Koguchi, and G. Kido, 
	\textit{unpublished}.
	
\bibitem{footnote1}
	In other words, large azimuthal asymmetry induces double-minima potential for carriers. Consequently, two localized levels are formed in QR c, leading to observation of the spectral doublet.
	
\bibitem{MB94} 
	J. Y. Marzin and G. Bastard, 
	Solid State Commun. \textbf{92}, 437 (1994). 
	
\bibitem{CH00} 
	M. Califano and P. Harrison, 
	Phys. Rev. B \textbf{61}, 10959 (2000). 

\bibitem{SGB99}
	O. Stier, M. Grundmann, and D. Bimberg, 
	Phys. Rev. B \textbf{59}, 5688 (1999).

\bibitem{SWK02}
	S. Sanguinetti, K. Watanabe, T. Kuroda, F. Minami, Y. Gotoh, 
	and N. Koguchi, 
	J. Crystal Growth \textbf{242}, 321 (2002).

\bibitem{MSG04}
	V.~Mantovani, S.~Sanguinetti, M.~Guzzi, E.~Grilli, M.~Grioli, 
	K.~Watanabe, and N.~Koguchi, J. Appl. Phys. \textbf{96}, 4416 (2004). 
	
	
\bibitem{PG94}
	L. Pavesi and M. Guzzi, 
	J. Appl. Phys. \textbf{75}, 4779 (1994). 
	
\bibitem{YSM04}
	M. Yamagiwa, N. Sumita, F. Minami, and N.~Koguchi, 
	J. Lumin. \textbf{108}, 379 (2004).

\bibitem{footnote2}
	The principal quantum number, $N$, characterizes both of the radial and vertical motions of carriers. However, we describe $N$ simply as the \textit{radial} quantum number, because the vertical component of electronic motion is essentially independent of $N (\leq 3)$. This is due to the strong confinement along the growth direction. See Fig.~\protect\ref{eh_level}(c). 
	
\bibitem{MK85}
	See, e.g., R. C. Miller and D. A. Kleinman, 
	J. Lumin. \textbf{30}, 520 (1985). 
			
\end{thebibliography}

\end{document}